\documentclass[aps,twocolumn,pra,showpacs]{revtex4}
\usepackage{amssymb}
\usepackage{amsmath}
\usepackage{graphicx}
\usepackage{subfigure}
\usepackage{amsfonts}

\begin{document}

\title{Quantum Optomechanics with Single Atom}
\author{Yue Chang, H. Ian, and C. P. Sun}
\email{suncp@itp.ac.cn}
\affiliation{Institute of Theoretical Physics, Chinese Academy of Sciences, Beijing,
100190, China}

\begin{abstract}
The recently increasing explorations for cavity optomechanical
coupling assisted by a single atom or an atomic ensemble have
opened an experimentally accessible fashion to interface quantum
optics and nano (micro) -mechanical systems. In this paper, we
study in details such composite quantum dynamics of photon, phonon
and atoms, specified by the triple coupling, which only exists in
this triple hybrid system: The cavity QED system with a movable
end mirror. We exactly diagonalize the Hamiltonian of the triple
hybrid system under the parametric resonance condition. We find
that, with the rotating-wave approximation, the hybrid system is
modeled by a generalized spin-orbit coupling where the orbital
angular momentum operator is defined through a Jordan-Schwinger
realization with two bosonic modes, corresponding to the mirror
oscillation and the single mode photon of the cavity. In the
quasi-classical limit of very large angular momentum, this system
will behave like a standard cavity-QED system described by the
Jaynes-Cummings model as the angular momentum operators are
transformed to bosonic operators of a single mode. We test this
observation with an experimentally accessible system with the atom
in the cavity with a moving mirror.
\end{abstract}

\maketitle

\section{Introduction}

Nowadays, there are an increasing number of researches on cavity
optomechanical systems assisted by atoms or atomic ensembles \cite%
{Tombesi,meystre,Ian}. Such hybrid systems show a convergence between
quantum optics and nano (micro) -mechanical systems. In this context, a
Fabry-Perot cavity is applied to form an approximate standing light wave
while one of the mirrors at the end of the cavity is allowed to oscillate.
Furthermore, the novel quantum natures are discovered in this composite
system by placing an atomic ensemble confined in a gas chamber inside the
cavity. With the helps of the atomic ensemble, theoretical explorations have
been made to show the possibilities to create not only the entanglement of
the cavity field and a macroscopical object \cite%
{Vitali,mancini,penrose,knight2}, i.e., a mirror, but also the entanglement
of atom-light-mirror \cite{Tombesi,Ian}. In this paper, we will consider
this atom assisted optomechanical (AAOPM) system with the strong coupling of
a single atom to the photon field inside the cavity, which is modified by
the moving end mirror of the cavity (see Fig. 1). We find that a three body
coupling term of photon, phonon and atom will play the crucial role in the
quantum dynamics of the triple system.

\begin{figure}[tbp]
\includegraphics[bb=17 520 351 789, width=8 cm, clip]{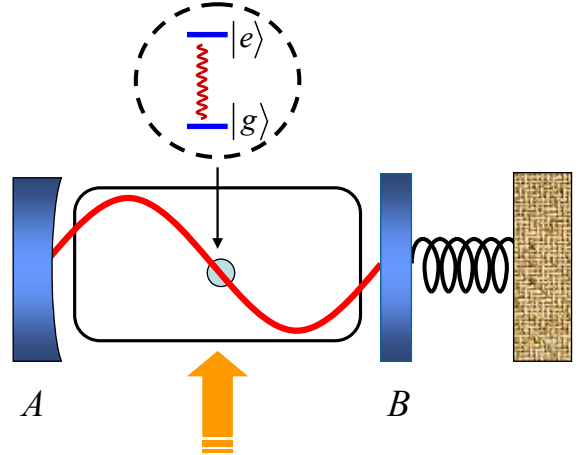}
\caption{\textit{The schematic of the atom assisted optomechanical
system. It contains an optical cavity ended with a fixed mirror A
and a slightly moving mirror B which is attached in a spring.
Inside the cavity there is a two level atom. }}
\end{figure}

The triple system we will refer to contains a two-level system interacting
with a single mode electromagnetic field inside the cavity with an
oscillating mirror. Due to the vibration of the cavity length, which is
conventionally thought to induce the light pressure term \cite%
{mancini2,mancini,ian2,meystre2}, we illustrate that the vibrating length
can also result in the triple coupling term of photon-atom-mirror. Under a
certain condition concerning the frequencies and the coupling strength in
the so-called parametric resonance, we can exactly diagonalize the
Hamiltonian to obtain the eigenstates and the eigenvalues. Here, the
eigenstate is the direct product of the mirror's state and the state of a
two body hybrid system, which consists of the photon and the atom, and can
be described by the Jaynes-Cummings (J-C) model. We consider how the
mirror's oscillation affects the cavity QED subsystem, by studying the
quantum decoherence \cite{zhang} of this subsystem. To this end, we
calculate the time evolution of the total system with the mirror initially
prepared in a coherent state. We use the decoherence factor (or the
Loschmidt echo (LE), the absolute square of the decoherence factor \cite%
{quan}), to characterize the influence of the mirror's motion, which is the
overlap of two wavefunctions of the mirror driven respectively by two
conditional Hamiltonians derived from two dressed states. We find that,
though the decoherence factor depends on the initial state of the mirror,
the LE has nothing to do with this initial coherent state. This means the
mirror's initial position only affect the decoherence factor in the form of
a phase factor.

The above condition for exact solution seems too special to be realized,
thus we consider a more general case with the rotating-wave approximation,
which only requires a set of matching frequencies rather than the coupling
strength, among the atom, the cavity field and the mirror. In this case, our
model is reduced to the generalized spin-orbit coupling model where the spin
is referred to as the internal energy level of the atom, while the orbit is
depicted by the quasi-orbital angular momentum defined by two bosonic modes
(the mirror oscillation and the single-mode photon of the cavity) through
the Jordan-Schwinger representation.

It is well known that the quantum system with a large angular momentum $L$
can be regarded as a classical rotor when the angular momentum approaches
infinity in the classical limit \cite{sun3}. This is because the component
variations $\Delta J_{x}$, $\Delta J_{y}$ and $\Delta J_{z}$ become
vanishingly small in comparison with the large angular momentum $L$. When
the angular momentum is large enough, but not infinite, the ladder operators
of any large angular momentum can behave as the creation and annihilation
operators of a single-mode boson \cite{jin,liu}. This point can be seen from
the Holstein-Primakoff transformation straightforwardly. This case is named
the quasi-classical limit and its significance in many-body physics can be
understood as the low-energy excitation above the ordered ground states.
This quasi-classical reduction of large angular momentum has been
extensively studied and applied in quantum storage \cite{sun,sun2,he}. Here,
we study this quasi-classical reduction by referring to the experimentally
accessible parameters in triple hybrid system. In this triple system, when
the frequency of the electromagnetic field almost matches the eigenfrequency
of the energy level spacing of the atom plus the mirror oscillation, the
photon and the mirror oscillation are coupled to form a composite object,
which exactly behaves as an angular momentum. Then our triple system is
modeled by the generalized spin-orbit coupling similar to the Hamiltonian
for the Paschen-Back effect without the radial-dependence \cite{landau}.
This observation means that the triple coupling system will be reduced to a
two-part coupling system described by the J-C model in cavity QED. At last,
we show this quasi-classical reduction indeed works well when the hybrid
excitation of the mirror plus photon is large enough.

The paper is organized as follows. In Sec. II, we model the AAOPM coupling
and reduce it to the generalized spin-orbit coupling under the frequency
matching condition. In Sec. III, we exactly solve the AAOPM model under the
parametric resonance condition and show the decoherence in Sec. IV. In Sec.
V, we compare the generalized spin-orbit coupling model to the J-C model
with their eigenstates and eigenvalues. Furthermore, in Sec. VI, we study
the dynamics of the generalized spin-orbit coupling model to demonstrate the
similarities and differences to that of the J-C model. In Sec. VII, we
summarize our results.

\section{Modeling the triple coupling of atom-photon-mirror}

In this section, we study an experimentally accessible AAOPM system, as
illustrated in Fig. 1. This system consists of three parts: an atom, the
photons inside a cavity, and a movable end mirror. We show that such a
hybrid system can be modeled by a spin-orbit coupling system, where the
orbital angular momenta are realized by the phonon of the mirror dressed by
the photon of the light field inside the cavity.

\begin{figure}[tbp]
\includegraphics[bb=19 569 278 787, width=6 cm, clip]{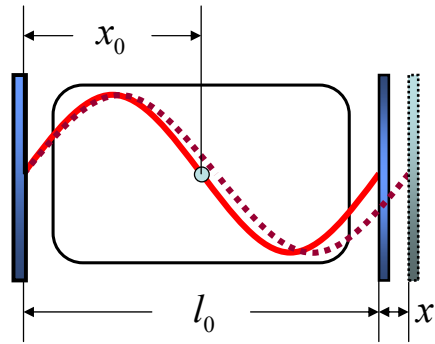}
\caption{\textit{Schematic of the origin of the triple coupling. When the
length of the cavity oscillates, besides the light pressure term, the
displacement of the mirror will affect the coupling strength between the
atom and the light. The the triple coupling term appears with this.}}
\end{figure}
The single mode electric field inside the cavity along the $x$-axis is
quantized as%
\begin{equation}
E(x_{0},t)=\varepsilon a\sin kx_{0}+h.c.\text{,}
\end{equation}%
where $x_{0}$ is the position of the atomic center of mass, and $\varepsilon
=\sqrt{\omega _{0}/\epsilon _{0}V}$. The frequency of the cavity field is
dependent of the cavity length, $\omega _{0}=k=2\pi /l_{0}$; $\epsilon _{0}$
is the dielectric constant in vacuum; $l_{0}$ and $V$ are the length and the
volume of the cavity, respectively. Here, we omit the polarization of the
field, but this will not affect our final conclusion for a practical system.
When the length of the cavity slightly changes from $l_{0}$ to $l_{0}+x$ due
to the mirror's displacement $x$ (see Fig. 2), the electric field becomes%
\begin{equation}
E(x_{0},t)\approx \varepsilon a\sin kx_{0}-\left( \eta ^{\prime
}xa+h.c.\right) \text{,}
\end{equation}%
where $\eta ^{\prime }=\left( \sin kx_{0}+kx_{0}\cos kx_{0}\right)
\varepsilon /l_{0}.$

Then, the total Hamiltonian reads%
\begin{eqnarray}
H &=&\omega _{0}a^{\dag }a+\omega _{M}b^{\dag }b-\xi \left( b+b^{\dag
}\right) a^{\dag }a+\omega _{e}S_{z}  \notag \\
&&+\left[ gaS_{+}+\eta \left( b+b^{\dag }\right) aS_{+}+h.c.\right] \text{.}
\label{a}
\end{eqnarray}%
where $a^{\dag }$ $\left( b^{\dag }\right) $ is the creation operator of the
oscillating mode of the cavity field (mirror); $S_{z}$, $S_{+}$ and $S_{-}$
are the spin operators, which represent the transitions among the atomic
inner states. $g=-\mu \varepsilon \sin kx_{0}$ is the
atomic-position-dependent coupling strength between the cavity field and the
atom, where $\mu $ is the electric-dipole transition matrix element. $\xi
=\omega _{0}/l_{0}\sqrt{2M\omega _{M}}$, where $\omega _{M}$ is the
frequency of the mirror's oscillation and $M$ is the mass of the mirror,
describes the light pressure, while $\eta =\eta ^{\prime }\mu /\sqrt{%
2M\omega _{M}}$ denotes the coupling strength of \textquotedblleft three
body\textquotedblright\ due to the vibration of the mirror. With some
experimentally feasible parameters, there exists the situation where $\eta $
is on the same order of magnitude of $\xi $, for which the strong coupling
region \cite{meystre3} is reached (e.g., $g=\omega _{0}=10^{15}Hz$ and $%
kx_{0}=\pi /2$, then $\eta =-\xi $). In this case, the triple coupling term%
\begin{equation}
V_{T}=\eta \left( b+b^{\dag }\right) a\left\vert e\right\rangle \left\langle
g\right\vert +h.c.
\end{equation}%
should not be neglected. In the following discussion, we will focus on this
strong coupling case.

It follows from the Hamiltonian in Eq. (\ref{a}) that, when $\sin
kx_{0}\rightarrow 0$, i.e., $g\rightarrow 0$, the J-C type interaction $%
gaS_{+}+h.c.$ vanishes, but the three-body interaction remains. Near the
photon-phonon resonance case where the frequencies satisfy $\omega
_{e}+\omega _{M}-\omega _{0}\approx 0$, the rotating-wave approximation
reduces the Hamiltonian to%
\begin{equation}
H_{\text{RWA}}=\omega _{0}a^{\dag }a+\omega _{M}b^{\dag }b+\omega
_{e}S_{z}+\eta \left( b^{\dag }aS_{+}+h.c.\right) \text{.}  \label{l}
\end{equation}%
In the following discussions, we invoke the Jordan-Schwinger representation
of the SO(3) group \cite{sakurai}%
\begin{equation}
L_{+}=a^{\dag }b\text{, }L_{-}=ab^{\dag }\text{, }L_{z}=\frac{1}{2}\left(
a^{\dag }a-b^{\dag }b\right) ,
\end{equation}%
where the commutation relations of the angular momenta%
\begin{equation}
\left[ L_{+},L_{-}\right] =2L_{z},\left[ L_{z},L_{+}\right] =L_{+},\left[
L_{z},L_{-}\right] =-L_{-}
\end{equation}%
are satisfied due to the generic commutation relations between the $a$ and $%
b $ bosons. Then, the Hamiltonian can be rewritten as%
\begin{equation}
H_{\text{RWA}}=\Omega N+\kappa L_{z}+\omega _{e}S_{z}+\eta \left(
L_{-}S_{+}+L_{+}S_{-}\right) ,
\end{equation}%
where $N=a^{\dag }a+b^{\dag }b$, $\Omega =\left( \omega _{0}+\omega
_{M}\right) /2$, and $\kappa =\omega _{0}-\omega _{M}$.

We remark that the three body interaction in the AAOPM system can be
approximately modeled with the x-y coupling%
\begin{equation}
\eta \left( L_{-}S_{+}+L_{+}S_{-}\right) =2\eta \left(
L_{x}S_{x}+L_{y}S_{y}\right) .
\end{equation}%
This is a kind of \textquotedblleft spin-orbit coupling\textquotedblright\
referred to as Paschen-Back effect \cite{landau}. The \textquotedblleft
orbital\textquotedblright\ angular momentum defined by $L_{\pm }$ and $L_{z}$
essentially results from the joint excitation of photon and phonon.
Physically, this excitation can be understood as a kind of effective
mechanical oscillation of the mirror, which is dressed by the single mode
photon. Such kinds of dressed bosons just satisfy the angular momentum
algebra.

\section{Exact solutions for parametric resonance}

From Eq. (\ref{a}), obviously the particle number operator $N=a^{\dag
}a+S_{z}$ is conserved, i.e., $[N,H]=0$. Then, in the subspace $\left\{
\left\vert n_{a}+1,g\right\rangle ,\text{ }\left\vert n_{a},e\right\rangle
\right\} $, where $\left\vert n_{a},g\left( e\right) \right\rangle $ denotes
that the photon is prepared in a Fock state $\left\vert n_{a}\right\rangle $
while the atom in the ground (excited) state. In the following we consider
the situation in which the eigen-equation of the Hamiltonian in Eq. (\ref{a}%
) can be solved exactly. To this end, we first show that the Hamiltonian is
formally expanded as follows:%
\begin{equation}
H_{n_{a}}=h\left( b,b^{\dag }\right) +H_{n_{a}}^{\prime },  \label{aa}
\end{equation}%
where%
\begin{equation}
h\left( b,b^{\dag }\right) =\omega _{e}+\omega _{0}n_{a}+\omega _{M}b^{\dag
}b-\xi n_{a}\left( b+b^{\dag }\right)
\end{equation}%
and%
\begin{equation}
H_{n_{a}}^{\prime }=\sqrt{n_{a}+1}\left(
\begin{array}{cc}
\frac{\Delta -\xi \left( b+b^{\dag }\right) }{\sqrt{n_{a}+1}} & g-\eta
\left( b+b^{\dag }\right) \\
g-\eta \left( b+b^{\dag }\right) & 0%
\end{array}%
\right) .  \label{bb}
\end{equation}%
In Eq. (\ref{bb}), $\Delta =\omega _{0}-\omega _{e}$ is the atom-photon
detuning. The above argument shows that in the subspace the total system can
be reduced to a spin-boson model defined by Eqs. (\ref{aa}-\ref{bb}).

Now let us temporarily leave the above concrete system to consider a more
general spin-boson model with the Hamiltonian%
\begin{equation}
H_{\text{SP}}=h\left( b,b^{\dag }\right) +W\left( S;b,b^{\dag }\right) ,
\end{equation}%
where $h\left( b,b^{\dag }\right) $ is a function that only depends on the
boson model and $W\left( S;b,b^{\dag }\right) $ is spin-dependent. There is
a seemingly trivial proposition for this spin-boson system: If the coupling
can be factorized as%
\begin{equation}
W\left( S;b,b^{\dag }\right) =f\left( b,b^{\dag }\right) M\left( S\right) ,
\label{cc}
\end{equation}%
with $f\left( b,b^{\dag }\right) $ ($M\left( S\right) $) depending on boson
(spin) only, then $H_{SP}$ can be exactly diagonalized through the
diagonalizations of the two pure boson systems with branch Hamiltonians%
\begin{equation}
H_{\text{SP}}^{\left( +,-\right) }=h\left( b,b^{\dag }\right) +\lambda
_{\left( +,-\right) }f\left( b,b^{\dag }\right) ,
\end{equation}%
where $\lambda _{+}$ and $\lambda _{-}$ are the eigenvalues of the C-number
coefficient matrix $M\left( S\right) $ in the basis $\left\{ \left\vert
+\right\rangle ,\text{ }\left\vert -\right\rangle \right\} $. The proof of
this proposition is rather straightforward and we only need to diagonalize $%
M\left( S\right) $ first.

Next we return to the concrete example.

We explore when $H_{n_{a}}^{\prime }$ can be factorized to $f\left(
b,b^{\dag }\right) M_{n_{a}}$, where $f\left( b,b^{\dag }\right) $ is a
function of $b$ and $b^{\dag }$, and $M_{n_{a}}$ is a $C$-number matrix.
Actually when the detuning $\Delta $ and the three coupling coefficients
have the relation
\begin{equation}
g\xi =\Delta \eta ,  \label{b}
\end{equation}%
which we call the parametric resonance, the Hamiltonian indeed becomes the
form of Eq. (\ref{cc}):%
\begin{equation}
H_{n_{a}}=h\left( b,b^{\dag }\right) +f\left( b,b^{\dag }\right) M_{n_{a}},
\end{equation}%
where%
\begin{equation}
f\left( b,b^{\dag }\right) =\Delta -\xi \left( b+b^{\dag }\right) ,
\end{equation}%
and%
\begin{equation}
M_{n_{a}}=\left(
\begin{array}{cc}
1 & \sqrt{n_{a}+1}\eta /\xi \\
\sqrt{n_{a}+1}\eta /\xi & 0%
\end{array}%
\right) .
\end{equation}

Thus, to diagonalize the Hamiltonian $H_{n_{a}}$, all we need to do is to
diagonalize the matrix $M_{n_{a}}$ and the left quadratic part formed by $b$
and $b^{\dag }$. Then, the eigenvalues of $H$ are obtained as%
\begin{eqnarray}
E_{j,n_{a},n_{b}} &=&\omega _{0}\left( n_{a}+\frac{1}{2}\right) +\frac{1}{2}%
\omega _{e}+\frac{\left( -\right) ^{j}}{2}R_{n_{a}}\Delta  \notag \\
&&+n_{b}\omega _{M}-\alpha _{jn_{a}}^{2}\text{,}  \label{c}
\end{eqnarray}%
for $j=1,2$, where%
\begin{equation}
R_{n_{a}}=\sqrt{1+\frac{4\eta ^{2}\left( n_{a}+1\right) }{\xi ^{2}}},
\end{equation}%
and%
\begin{equation}
\alpha _{jn_{a}}=\frac{\xi }{2\omega _{M}}\left( 2n_{a}+\left( -\right)
^{j}R_{n_{a}}+1\right) .
\end{equation}%
Here, $n_{a}$ $\left( n_{b}\right) $ represents the quantum number of the
photons (phonons). Correspondingly, the eigenstates of the AAOPM system are $%
\left\vert jn_{a}\right\rangle \otimes \left\vert n_{b}\right\rangle
_{_{_{jn_{a}}}}$, where the photon dressed state%
\begin{equation}
\left\vert 1n_{a}\right\rangle =\cos \theta _{n_{a}}\left\vert
n_{a}+1,g\right\rangle +\sin \theta _{n_{a}}\left\vert n_{a},e\right\rangle ,
\label{d}
\end{equation}%
and%
\begin{equation}
\left\vert 2n_{a}\right\rangle =-\sin \theta _{n_{a}}\left\vert
n_{a}+1,g\right\rangle +\cos \theta _{n_{a}}\left\vert n_{a},e\right\rangle ,
\label{e}
\end{equation}%
are defined by the mixing angle $\theta _{n_{a}}$:%
\begin{equation}
\tan \theta _{n_{a}}=\frac{2\eta \sqrt{n_{a}+1}}{\xi -\xi R_{n_{a}}},
\end{equation}%
while the mirror's states $\left\vert n_{b}\right\rangle _{_{_{jn_{a}}}}$ are%
\begin{equation}
\left\vert n_{b}\right\rangle _{_{_{jn_{a}}}}=\frac{1}{\sqrt{n_{b}!}}\left(
b^{\dag }-\alpha _{jn_{a}}\right) ^{n_{b}}D_{b}\left( \alpha
_{jn_{a}}\right) \left\vert 0\right\rangle ,
\end{equation}%
with the displacement operator $D_{b}\left( \alpha _{jn_{a}}\right) =\exp %
\left[ \alpha _{jn_{a}}\left( b^{\dag }-b\right) \right] .$

If $\xi =\eta =0$, the atom-assisted optomechanical coupling model goes back
to the J-C model, and the eigenvalues, together with the eigenstates,
degenerate to that of the J-C model. However, due to the vibration of the
mirror, there emerge fruitful results in our model due to the complex
three-body coupling. First, we examine the realization of the parametric
resonance condition, (\ref{b}), in experiments. Substituting the
experimental feasible parameters into the parametric resonance condition, we
know that it is easily satisfied if the detuning $\Delta $ is adjusted
properly to adapt to different positions of the atom. In experiments, $\xi $
and $\eta $ can reach $10^{5}$Hz, while $g$ is on the order of $10^{15}$Hz,
thus $\Delta $ can be on any order that lies on the atomic position. In the
special case when $\sin kx_{0}=0$, i.e., $g=0$, $\Delta =0$ is sufficient to
meet the condition in Eq. (\ref{b}).

It is observed from Eq. (\ref{c}) that the terms in the second line
obviously differ from the eigenvalues of the J-C model. The first term is
the mirror's eigenvalue, and the second one (without the sign) is expanded as%
\begin{equation}
\frac{\xi ^{2}n_{a}^{2}}{\omega _{M}}+\frac{\xi ^{2}}{4\omega _{M}}\left(
4n_{a}+\left( -\right) ^{j}R_{n_{a}}+1\right) \left( \left( -\right)
^{j}R_{n_{a}}+1\right) ,
\end{equation}%
where the first term $\xi ^{2}n_{a}^{2}/\omega _{M}$ describes the energy of
the light pressure term $\xi \left( b+b^{\dag }\right) a^{\dag }a$ \cite%
{knight2,knight}, but the left terms are induced by the atom-assisted
optomechanical coupling.

\section{Conditional dynamics for decoherence}

In this section we demonstrate the parametric resonance will lead to a
conditional dynamics with respect to two superpositions of atomic inner
states $\left\vert +\right\rangle $ and $\left\vert -\right\rangle $, which
is described by a non-demolition Hamiltonian.

From the above argument about the exact solvability of the AAOPM system, we
can find that the operator-valued Hamiltonian matrix%
\begin{equation}
H=\left[
\begin{array}{cc}
H_{SP}^{\left( +\right) } & 0 \\
0 & H_{SP}^{\left( -\right) }%
\end{array}%
\right]
\end{equation}%
is diagonalized with respect to the basis $\left\{ \left\vert +\right\rangle
,\text{ }\left\vert -\right\rangle \right\} $ for the spin part. Obviously
this is a non-demolition Hamiltonian with respect to the basis vectors $%
\left\vert +\right\rangle $ and $\left\vert -\right\rangle ,$and thus
results in the corresponding decoherence.

Driven by this non-demolition Hamiltonian, the factorized initial state for
the cavity QED system%
\begin{equation}
\left\vert \psi \left( 0\right) \right\rangle =\left( C_{+}\left\vert
+\right\rangle +C_{-}\left\vert -\right\rangle \right) \otimes \left\vert
\varphi \right\rangle
\end{equation}%
will evolve into an entanglement state%
\begin{equation}
\left\vert \psi \left( t\right) \right\rangle =C_{+}\left\vert
+\right\rangle \left\vert \varphi _{+}\left( t\right) \right\rangle
+C_{-}\left\vert -\right\rangle \left\vert \varphi _{-}\left( t\right)
\right\rangle ,
\end{equation}%
where%
\begin{equation*}
\left\vert \varphi _{\pm }\left( t\right) \right\rangle =e^{-iH_{SP}^{\left(
\pm \right) }t}\left\vert \varphi \right\rangle ,
\end{equation*}%
and the extent of decoherence due to this quantum entanglement is
characterized by the so-called decoherence factor%
\begin{eqnarray}
D\left( t\right) &=&\langle \varphi _{+}\left( t\right) \left\vert \varphi
_{-}\left( t\right) \right\rangle  \notag \\
&=&Tr\left( \rho \left( 0\right) e^{iH_{SP}^{\left( +\right)
}t}e^{-iH_{SP}^{\left( -\right) }t}\right) ,
\end{eqnarray}%
and its norm square $L\left( t\right) =\left\vert D\left( t\right)
\right\vert ^{2}$ is the so-called Loschmidt echo \cite{quan}.

In the context of quantum chaos, the Loschmidt echo characterizes the
sensitivity of evolution of quantum system in comparison with the butterfly
effect in classical chaos: Starting from the same initial state, the quantum
system is separately driven by two slight different Hamiltonians. Quantum
chaos is implied by the much larger differences in the two corresponding
final states; namely, their overlap (Loschmidt echo) vanishes to illustrate
the dynamical sensitivity of the quantum chaos system.

Next we return to the concrete system.

Consider the time evolution of the system when the initial state is as
follows:%
\begin{equation}
\psi \left( 0\right) =\sum_{j,n_{a}}\lambda _{jn_{a}}\left\vert
jn_{a}\right\rangle \otimes \left\vert \beta \right\rangle _{b},
\end{equation}%
where $\left\vert \beta \right\rangle _{b}$ is a coherent state of phonon
that satisfies%
\begin{equation}
b\left\vert \beta \right\rangle _{b}=\beta \left\vert \beta \right\rangle
_{b},
\end{equation}%
$\left\vert jn_{a}\right\rangle $, $j=1,2$, is the dressed state mentioned
in last section, and $\lambda _{jn_{a}}$ is the weight of each dressed
state. Therefore, at time $t$ the wave function of the total system is%
\begin{equation}
\psi \left( t\right) =\sum_{j,n_{a}}\lambda _{jn_{a}}\left\vert
jn_{a}\right\rangle \otimes e^{-jC_{_{jn_{a}}}t}\left\vert \left( \beta
-\alpha _{jn_{a}}\right) e^{-j\omega _{M}t}+\alpha _{jn_{a}}\right\rangle
_{b}
\end{equation}%
where%
\begin{eqnarray}
C_{_{jn_{a}}} &=&\frac{1}{2}\omega _{e}+\omega _{0}\left( n_{a}+\frac{1}{2}%
\right) +\frac{\left( -\right) ^{j}}{2}R_{n_{a}}\Delta  \notag \\
&&-\frac{\xi ^{2}}{4\omega _{M}}\left( 2n_{a}+\left( -\right)
^{j}R_{n_{a}}+1\right) ^{2}.
\end{eqnarray}

The mirror's motion will result in the collapse of the decoherence, with the
Loschmidt echo (LE) being%
\begin{eqnarray}
LE_{_{n_{a},m_{a}}}^{j_{1},j_{2}} &=&\left\vert _{b}\left\langle \beta
\right\vert e^{iH_{j_{2}n_{a}}t}e^{-iH_{j_{1}m_{a}}t}\left\vert \beta
\right\rangle _{b}\right\vert ^{2}  \notag \\
&=&\exp \left[ 2\left( \Delta _{n_{a},m_{a}}^{j_{1},j_{2}}\right) ^{2}\left(
\cos \omega _{M}t-1\right) \right] ,  \label{f}
\end{eqnarray}%
where%
\begin{equation}
\Delta _{n_{a},m_{a}}^{j_{1},j_{2}}=\alpha _{in_{a}}-\alpha _{jm_{a}}.
\end{equation}%
We notice that $\beta $ does not play any role in the LE. Note that the
mirror's initial coherent state evolves to another coherent state, and $%
\beta $ only determine the initial position of the center of the wavepacket $%
\langle x\left\vert \beta \right\rangle _{b}$. Thus the overlap of the two
wavepackets $\exp (-iH_{j_{1}n_{a}}t)\left\vert \beta \right\rangle _{b}$
and $\exp (-iH_{j_{2}m_{a}}t)\left\vert \beta \right\rangle _{b}$ is
independent of $\beta $. Physically, this fact shows that the decoherence of
the cavity-QED system is irrelevant to the phonon excitations of the mirror
if its wavepacket is Gaussian.

In Fig. 3, for different photon-phonon excitations, we plot the time
evolution of the Loschmidt echo with the parameters set to $\omega
_{0}=10^{15}$Hz, $\eta =10\omega _{0}/l\sqrt{2M\omega _{M}}=10\xi $, $%
l_{0}=1\mu $m, $M=10^{-10}$kg, $\omega _{M}=10^{9}$Hz. From Fig. 3, we see
that all the curves have the same period, i.e., $2\pi /\omega _{M}$, and the
larger the difference of the photon number $\left\vert
n_{a}-m_{a}\right\vert $ is, the larger the amplitudes of the curve are. We
remark that the large photon number means a classical electromagnetic field,
and thus the quantum decoherence of the atomic inner states reflects the
classical transition of optical field from the quantum regime.

\begin{figure}[tbp]
\includegraphics[bb=0 0 217 123, width=9 cm, clip]{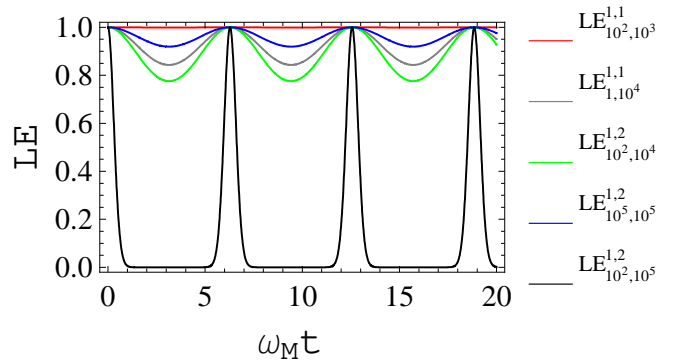}
\caption{\textit{The time evolution of the Loschmidt echo with different
indexes of LE as shown in the equation above, which corresponds to
photon-phonon excitations.}}
\end{figure}

\section{Generalized spin-orbit coupling in comparison with Jaynes-Cummings
model}

In Sec. II, we have derived the generalized spin-orbit coupling model under
the rotating-wave approximation, which can be rewritten as%
\begin{equation}
H_{RWA}=\Omega N+\kappa L_{z}+\omega _{e}S_{z}+2\eta \left( \vec{L}\cdot
\vec{S}-L_{z}S_{z}\right) ,  \label{g}
\end{equation}%
Here, $\kappa $ and $\omega _{e}$ characterize the coupling of the angular
momentum and the spin to the external field, respectively; $2\eta $ is the
coupling strength of the spin-orbit.

This model can be studied by exactly diagonalizing the model Hamiltonian in
Eq. (\ref{g}) within its invariant subspace spanned by%
\begin{equation}
\left\vert l,m,e\right\rangle =\left\vert l,m\right\rangle \otimes
\left\vert \uparrow \right\rangle
\end{equation}%
and%
\begin{equation}
\left\vert l,m+1,g\right\rangle =\left\vert l,m+1\right\rangle \otimes
\left\vert \downarrow \right\rangle .
\end{equation}%
Here, $\left\vert l,m\right\rangle $ is the standard angular momentum basis,
while $\left\vert \uparrow \right\rangle $ and $\left\vert \downarrow
\right\rangle $ denote the spin-up and spin-down vectors, respectively. In
this basis, the spin-orbit coupling Hamiltonian in Eq. (\ref{g}) is reduced
to a quasi-diagonal matrix with an $2\times 2$ block%
\begin{equation}
H_{_{RWA}}^{l}=2\Omega l+\left( m+\frac{1}{2}\right) \delta +\frac{1}{2}%
\left(
\begin{array}{cc}
\delta & g_{lm} \\
g_{lm} & -\delta%
\end{array}%
\right) ,
\end{equation}%
where $\Delta =\omega _{e}-\delta ,$and%
\begin{equation}
g_{lm}=2\eta \sqrt{l\left( l+1\right) -m\left( m+1\right) }.
\end{equation}%
Then it can be further diagonalized to obtain the eigenvectors%
\begin{equation}
\left\vert +lm\right\rangle =\cos \theta _{lm}\left\vert l,m,e\right\rangle
+\sin \theta _{lm}\left\vert l,m+1,g\right\rangle ,
\end{equation}%
and%
\begin{equation}
\left\vert -lm\right\rangle =-\sin \theta _{lm}\left\vert l,m,e\right\rangle
+\cos \theta _{lm}\left\vert l,m+1,g\right\rangle ,
\end{equation}%
with the corresponding eigenvalues%
\begin{equation}
E_{\pm lm}=2\Omega l+\left( m+\frac{1}{2}\right) \kappa \pm \frac{1}{2}%
R_{lm},
\end{equation}%
where $\tan \theta _{lm}=g_{lm}/\left( \delta +R_{lm}\right) $ and $R_{lm}=%
\sqrt{g_{lm}^{2}+\delta ^{2}}.$

From the spectrum structure of the AAOPM system described above, we
demonstrate that the triple coupling system can realize the entanglement
between an orbital angular momentum and a spin. The $z$-component of the
total angular momentum is conserved. Thus, while the orbital angular
momentum is flipped from down (up) to up (down), the spin will make a
reverse flip.

Now we can consider the quasi-classical limit of the above spin-orbit
coupling model for $l$ large enough with low excitation. Obviously, the
above basis vector in this limit becomes a Fock state, i.e, $\left\vert
l,m\right\rangle \rightarrow \left\vert n\right\rangle ,$ where $n=l+m,$ $%
n/l\ll 1$ ; while $\tan \theta _{lm}\rightarrow \tan \theta
_{n}=g_{n}/\left( \delta +R_{n}\right) $. Correspondingly, the eigenstates
become the dressed states in the usual J-C model with the eigenvalues%
\begin{equation}
E_{\pm lm}\rightarrow E_{\pm n}=\left( n+\frac{1}{2}\right) \kappa \pm \frac{%
1}{2}R_{n},\text{ }i=1,2,
\end{equation}%
where $g_{n}=2\eta \sqrt{2l\left( n+1\right) }$ and $R_{n}=\sqrt{%
g_{n}^{2}+\delta ^{2}}.$

We remark that the system we considered in the limit above can also be
described by the J-C model%
\begin{equation}
H_{J-C}=\kappa a^{\dag }a+\omega _{e}S_{z}+\eta \sqrt{2l}\left(
aS_{+}+a^{\dag }S_{-}\right) .
\end{equation}%
The correspondence between the Fock state $\left\vert n\right\rangle $ and
the standard angular momentum basis is%
\begin{equation}
\left\vert n\right\rangle \Leftrightarrow \left\vert l,l-n\right\rangle .
\end{equation}

Actually, the above equivalence of spin-orbit coupling model and J-C
interaction can be found directly by considering the Holstein-Primakoff
mapping%
\begin{equation}
L_{+}=a^{\dag }\sqrt{2l-a^{\dag }a},L_{-}=L_{+}^{\dag },L_{z}=a^{\dag }a-l
\end{equation}%
in the large $l$ limit.

Next we come back to the practical physics of the AAOPM system. Then the
joint states $\left\vert l,m,e\left( g\right) \right\rangle =\left\vert
l,m\right\rangle \otimes \left\vert \uparrow \left( \downarrow \right)
\right\rangle $ is re-expressed in terms of the two Fock states $\left\vert
l+m\right\rangle _{a}$ and $\left\vert l-m\right\rangle _{b}$ as%
\begin{equation}
\left\vert l,m,e\left( g\right) \right\rangle =\left\vert l+m\right\rangle
_{a}\left\vert l-m\right\rangle _{b}\left\vert e\left( g\right)
\right\rangle .
\end{equation}%
Here $\left\vert l+m\right\rangle _{a}$ and $\left\vert l-m\right\rangle
_{b} $ represent the states with $l+m$ photons and $l-m$ phonons of the
mirror's vibration mode respectively.

In Fig. 4 and Fig. 5, with different values of $l$ and $m$, we plot $E_{\pm
lm}$ as the functions of $\omega _{e}$. Here, we take the physical
parameters as $\omega _{0}=1.9\times 10^{15}$Hz, $\eta =\omega _{0}/l\sqrt{%
2M\omega _{M}}$, $l_{0}=1\mu $m, $M=10^{-10}$kg, $\omega _{M}=10^{9}$Hz.
\begin{figure}[tbp]
\includegraphics[bb=0 0 205 128, width=8 cm, clip]{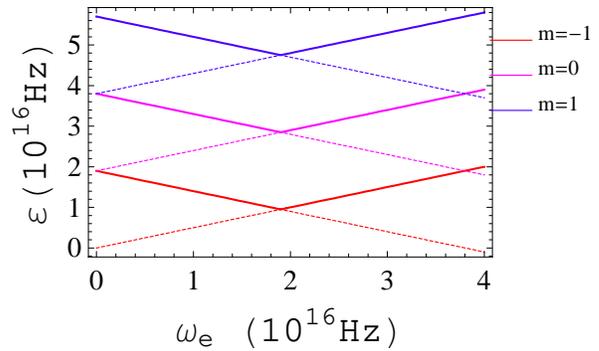}
\caption{\textit{Schematic of the relation of the eigenvalues of the
Atom-Photon-Mirror coupling system with the spacing frequency of the
two-level atom. Here we take m=-1, 0, 1 when l=1. Each real line represents $%
\protect\varepsilon _{+lm}$ while the dashed with the same color represents $%
\protect\varepsilon _{+lm}$.}}
\end{figure}

\begin{figure}[tbp]
\includegraphics[bb=0 0 217 127, width=8 cm, clip]{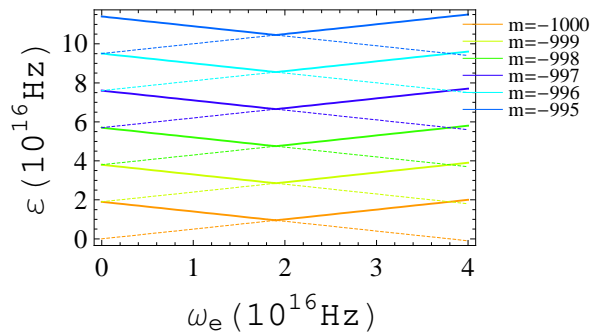}
\caption{\textit{Schematic of the lowest 5 levels when l=1000.}}
\end{figure}
As shown in the figures 4 and 5, when $l$ is fixed, the spectrum diagram of
the AAOPM system looks quite like that of the J-C model's \cite{orszag}.
However, in general, its number of energy levels are much more than that in
the J-C model's within the same energy range. Furthermore, we can see from
the several lowest levels that there exist small differences under different
values of $l$, such as $E_{+,1,-1}$ vs. $E_{+,1000,-1000}$ , $E_{-,1,0}$ vs.
$E_{-,1000,-999}$ and so on. Accordingly, the larger the value of $l$ (the
orbital angular momentum) is, the closer the spectrum is to the
corresponding ones in the J-C model. For evidence, we have plotted the
curves with a much more larger range of the variable $\omega _{e}$ that is
not valid in our rotating wave approximation that requires $\left\vert
\omega _{e}-\omega _{0}\right\vert \ll \omega _{M}$.

\section{Quasi-classical dynamics via Jaynes-Cummings Model}

In the above sections, we have shown the similarity between the triple
hybrid system and the J-C model in their energy spectra. Now we continue to
consider this similarity in quantum dynamics, which is referred to as the
so-called quasi-classical one as we make the analysis for very large angular
momentum.

We consider generally the system described by the Hamiltonian Eq. (\ref{g})
is initially prepared in the state represented by the density matrix
\begin{equation}
\rho \left( 0\right) =\sum_{ijkl}\lambda _{ij;kl}\left\vert \psi _{ij}\left(
0\right) \right\rangle \left\langle \psi _{kl}\left( 0\right) \right\vert ,
\end{equation}%
where the joint states $\left\vert \psi _{ij}\left( 0\right) \right\rangle
=\left\vert i\right\rangle _{a}\left\vert j\right\rangle _{b}\left\vert
e\right\rangle $ denote the initial factorized structure of the triple
system. According to the similarity between the generalized L-S coupling
system and the triple coupling system we mentioned in the last section, the
density matrix for the time evolution reads%
\begin{equation}
\rho \left( t\right) =\sum_{ijkl}\lambda _{ij;kl}\left\vert \psi _{ij}\left(
t\right) \right\rangle \left\langle \psi _{kl}\left( t\right) \right\vert ,
\end{equation}%
where the branch wavefunctions%
\begin{eqnarray}
\left\vert \psi _{ij}\left( t\right) \right\rangle &=&e^{-i\omega
_{ij}t}[\lambda _{ij}^{e}\left( t\right) \left\vert i\right\rangle
_{a}\left\vert j\right\rangle _{b}\left\vert e\right\rangle  \notag \\
&&-\lambda _{i+1,j-1}^{g}\left( t\right) \left\vert i+1\right\rangle
_{a}\left\vert j-1\right\rangle _{b}\left\vert g\right\rangle ],  \label{h}
\end{eqnarray}%
for $\omega _{ij}=\omega _{0}i+\omega _{M}j+\delta /2$, are determined by
the time dependent parameters defined by%
\begin{equation}
\lambda _{ij}^{e}\left( t\right) =\cos \frac{\Omega _{ij}t}{2}-\mathrm{i}%
\cos 2\phi _{ij}\sin \frac{\Omega _{ij}t}{2},
\end{equation}%
\begin{equation}
\lambda _{i+1,j-1}^{g}\left( t\right) =\mathrm{i}\sin 2\phi _{ij}\sin \frac{%
\Omega _{ij}t}{2},
\end{equation}%
\begin{equation}
\tan \phi _{ij}=\frac{2\eta \sqrt{j\left( i+1\right) }}{\omega _{e}-\delta
+\Omega _{ij}},
\end{equation}%
\begin{equation}
\Omega _{ij}=\sqrt{4\eta ^{2}\left[ j\left( i+1\right) \right] +\left(
\omega _{e}-\delta \right) ^{2}}\text{.}  \label{i}
\end{equation}

It is noticed that the above Eqs. (\ref{h}-\ref{i}) have forms similar to
that for the J-C model in the limit that $i/j\ll 1$. We remark that this
limit means a low excitation of the joint system consisting of the photon
and phonon, i.e., there are only a few photons stimulated so that the joint
system is almost prepared in the lowest weight state $\left\vert
l,m=-l\right\rangle $. The above equations give us rich information on the
complex system, from which all the physically relevant quantities to the
cavity field, the atom and the mirror can be obtained.\ In Eq. (\ref{h}), $%
\left\vert \lambda _{ij}^{e}\left( t\right) \right\vert ^{2}$ ($\left\vert
\lambda _{ij}^{g}\left( t\right) \right\vert ^{2}$) is proportional to the
probability that at time $t$, there are $i$ photons, $j$ quanta of the
mirror's vibration mode and one atom in the excited (ground) state.
Therefore the probability $p_{n}\left( t\right) $ that $n$ photons are
measured is%
\begin{eqnarray}
p_{n}\left( t\right) &=&\sum_{j}\lambda _{nj;nj}\left( c_{_{nj}}^{2}\left(
t\right) +\cos ^{2}2\phi _{nj}s_{_{nj}}^{2}\left( t\right) \right)  \notag \\
&&+\left\vert \lambda _{n-1,j;n-1,j}\right\vert ^{2}\sin ^{2}2\phi
_{n-1,j}s_{_{n-1,j}}^{2}\left( t\right) ,  \label{j}
\end{eqnarray}%
where $c_{_{nj}}\left( t\right) =\cos \Omega _{nj}t/2$, and $s_{_{nj}}\left(
t\right) =\sin \Omega _{nj}t/2$\

Another important quantity is the population inversion $W\left( t\right) $
depending on the probability amplitudes $\lambda _{ij}^{e}\left( t\right) $
and $\lambda _{ij}^{g}\left( t\right) $ as%
\begin{eqnarray}
W\left( t\right) &=&\sum_{i,j}\lambda _{ij;ij}\left( \left\vert \lambda
_{ij}^{e}\left( t\right) \right\vert ^{2}-\left\vert \lambda
_{i+1,j-1}^{g}\left( t\right) \right\vert ^{2}\right)  \notag \\
&=&\sum_{i,j}\lambda _{ij;ij}[c_{_{ij}}^{2}\left( t\right) +\cos 4\phi
_{ij}s_{_{ij}}^{2}\left( t\right) ].  \label{k}
\end{eqnarray}

Note that if the initial state of the mirror is the vacuum state, i.e., $%
\lambda _{ij;kl}\propto \delta _{j0}\delta _{l0}$, then it follows from Eqs.
(\ref{j}) and (\ref{k}) that $p_{n}\left( t\right) =\sum_{j}\lambda _{nj;nj}$
and $W\left( t\right) =1,$ both of which are time independent no matter
which state the light field is initially in. This result can be explained as
follows: with the rotating-wave approximation, we only hold the slowly
varying terms in the original Hamiltonian to obtain effective Hamiltonian (%
\ref{l}). These terms make a transition from the atom's upper level state $%
\left\vert e\right\rangle $ to the ground state $\left\vert g\right\rangle $%
, together with a decrement of the quanta of the mirror's vibration mode and
an increment of the photon number, or vice versa. Thus when initially the
mirror is in vacuum and the atom is in the excited state, the total state
can not evolve to the \textquotedblleft dressed state\textquotedblright ,
but only stays in the initial state companied by a dynamical phase factor.

The above obtained results are very similar to that of the J-C model: $%
p_{n}\left( t\right) $ and $W\left( t\right) $ also contain many Rabi
oscillations with various frequencies, and in different initial states, $%
p_{n}\left( t\right) $ and $W\left( t\right) $ behave differently.

Next we consider that the mirror is initially in a thermal state, while the
photon field is in one of several different states: a thermal state, a Fock
state $\left\vert n_{0}\right\rangle _{a}$ and a coherent state $\left\vert
\alpha \right\rangle _{a}$. We obtain different initial parameters as
follows:%
\begin{equation}
\lambda _{ij;ij}^{thermal}=\frac{e^{\beta \omega _{0}}-1}{e^{\beta \omega
_{0}\left( i+1\right) }}\frac{e^{\beta \omega _{M}}-1}{e^{\beta \omega
_{M}\left( j+1\right) }},
\end{equation}%
\begin{equation}
\lambda _{ij;ij}^{Fock}=\delta _{in_{0}}\frac{e^{\beta \omega _{M}}-1}{%
e^{\beta \omega _{M}\left( j+1\right) }},
\end{equation}%
\begin{equation}
\lambda _{ij;ij}^{coherent}=\exp \left( -\left\vert \alpha \right\vert
^{2}\right) \frac{\left\vert \alpha \right\vert ^{2i}}{i!}\frac{e^{\beta
\omega _{M}}-1}{e^{\beta \omega _{M}\left( j+1\right) }},
\end{equation}%
where $\beta =1/k_{B}T$, $T$ is the temperature.

In Fig. 6, 7 and 8, we plot the evolution of $W\left( t\right) $ in the
three initial states mentioned above with the parameters $T=1$K, $M=10^{-10}$%
kg, $l_{0}=1\mu $m, $\omega _{0}=\eta =1.9\times 10^{15}$Hz, $\omega
_{e}=\omega _{0}-0.999\omega _{M}$, $\omega _{M}=10^{9}$Hz, $n_{0}=10$, $%
\alpha =10$. It can be seen from the figures that in each case, as time
increases, collapses and revivals appear cyclically, but the time duration
in which each collapse and each revival take place differs from each other
because of different $\lambda _{ij;ij}$ that represents the weight of the
Rabi oscillation with fixed frequency. This behavior of collapse and revival
of inversion is repeated with increasing time, with the amplitude of Rabi
oscillations decreased and the time duration in which the revival takes
place increased and ultimately overlapping with the earlier revival.

Note that the temperature is so low that $\left[ \exp \left( \beta \omega
_{0}\right) -1\right] /\exp \left[ \beta \omega _{0}\left( i+1\right) \right]
\rightarrow \delta _{i0}$, and thus the case described in Fig. 6 reflects
the phenomenon that the atomic transition between the upper and the lower
levels can happen even when the light field is initially prepared in the
vacuum state. This is obviously a purely quantum effect to prove the role of
vacuum. In Fig. 7 we observe that even if the cavity field in a Fock state,
the collapse and revival appear explicitly. This case differs from the J-C
model based collapse and revival phenomenon, in which the evolution of
inversion is just a cosine curve when the field in a Fock state.

\begin{figure}[tbp]
\includegraphics[bb=7 418 587 820, width=8 cm, clip]{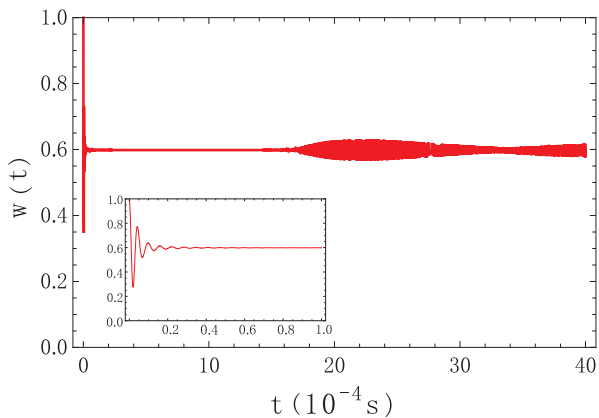}
\caption{\textit{Time evolution of the population inversion $W\left(
t\right) $ for an initially thermal state for both the cavity field and the
mirror.}}
\end{figure}

\begin{figure}[tbp]
\includegraphics[bb=12 420 593 802, width=8 cm, clip]{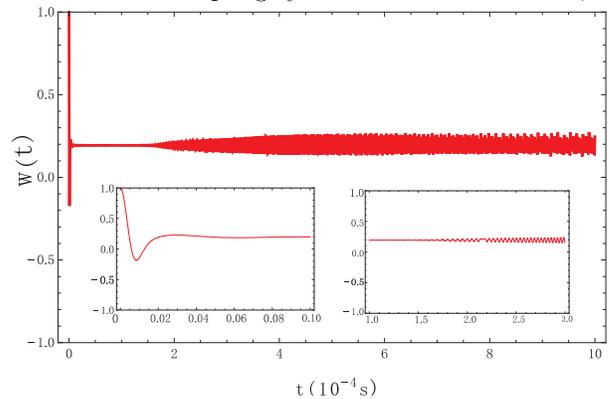}
\caption{\textit{Time evolution of the population inversion $W\left(
t\right) $ for an initially thermal state for mirror and a Fock state $%
\left\vert n_{0}\right\rangle _{a}$ for the field with $n_{0}=10$.}}
\end{figure}

\begin{figure}[tbp]
\includegraphics[bb=12 500 530 828, width=8 cm, clip]{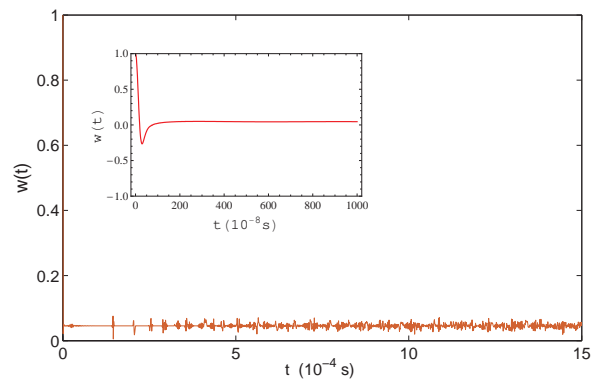}
\caption{\textit{Time evolution of the population inversion $W\left(
t\right) $ for an initially thermal state for mirror and a coherent state $%
\left\vert \protect\alpha \right\rangle _{a}$ for the field with $\protect%
\alpha =10$.}}
\end{figure}

\section{Conclusion and Remarks}

We have shown a AAOPM coupling in the triple hybrid system composing of
atoms, a cavity field and a movable mirror and discovered that under
parametric resonance condition, this complicated model can be solved
exactly. Furthermore, we have demonstrated that this triple hybrid system
can be modeled by generalized L-S coupling under the rotating-wave
approximation. It is shown that the composite object formed by the\
cavity-field-dressed mirror acts like an orbital angular momentum. Then we
studied the physically intrinsic relation between the generalized L-S
coupling system and the J-C model, i.e., when the orbital angular momentum
is large enough, the former is quite like the latter. Same to the
generalized L-S coupling system, in quasi-classical limit, the ladder
operators behave as the bosonic operators and thus the large angular
momentum can be regarded as \textquotedblleft excitons\textquotedblright\ in
the low excitation limit. We also investigated some characteristic
properties of the J-C model in our triple hybrid system and discovered their
similarities and differences.

When preparing this paper we find a paper \cite{x x yi}, where the triple
interaction was introduced.

\acknowledgments C. P. Sun acknowledges supports by the NSFC with Grants No.
10474104, No. 60433050, and No. 10704023, NFRPCNo. 2006CB921205 and
2005CB724508.



\begin{thebibliography}{99}
\bibitem{Tombesi} C. Genes, D. Vitali, and P. Tombesi, Phys. Rev. A \textbf{%
77}, 050307 (2008).

\bibitem{meystre} D. Meiser and P. Meystre, Phys. Rev. A \textbf{73}, 033417
(2006).

\bibitem{Ian} H. Ian, Z. R. Gong, Yu-xi Liu, C. P. Sun and Franco Nori,
Phys. Rev. A \textbf{78}, 013824 (2008).

\bibitem{Vitali} D. Vitali, S. Gigan, A. Ferreira, H. R. B\"{o}hm, P.
Tombesi, A. Guerreiro, V. Vedral, A. Zeilinger, and M. Aspelmeyer, Phys.
Rev. Lett. 98, 030405 (2007).

\bibitem{penrose} W. Marshall, C. Simon, R. Penrose, and D. Bouwmeester,
Phys. Rev. Lett. \textbf{91}, 130401 (2003).

\bibitem{knight2} S. Bose, K. Jacobs, and P. L. Knight, Phys. Rev. A 59,
3204 - 3210 (1999).

\bibitem{mancini} S. Mancini and P. Tombesi, Phys. Rev. A 49, 4055 (1994).

\bibitem{mancini2} S. Mancini1, V. Giovannetti, D. Vitali, and P. Tombesi,
Phys. Rev. Lett. 88, 120401 (2002).

\bibitem{ian2} Z. R. Gong, H. Ian, Yu-xi Liu, C. P. Sun, Franco Nori,
arXiv:0805.4102 (2008).

\bibitem{meystre2} M. Bhattacharya and P. Meystre, Phys. Rev. Lett. 99,
073601 (2007).

\bibitem{zhang} P. Zhang, X. F. Liu, and C. P. Sun , Phys. Rev. A 66, 042104
(2002).

\bibitem{sun3} C. P. Sun, Phys. Rev. A 48, 898 (1993).

\bibitem{jin} Y. X. Liu, N. Imoto, \c{S}. K. \"{o}zdemir, G. R. Jin, and C.
P. Sun, Phys. Rev. A 65, 023805 (2002).

\bibitem{liu} G. R. Jin, P. Zhang, Yu-xi Liu, and C. P. Sun, Phys. Rev. B
68, 134301 (2003).

\bibitem{sun} C. P. Sun, Y. Li and X. F. Liu, Phys. Rev. Lett. \textbf{91},
147903 (2003).

\bibitem{sun2} Y. Li and C. P. Sun, Phys. Rev. A \textbf{69}, 051802 (2004).

\bibitem{he} L. He, Y. X. Liu, S. Yi, C. P. Sun, and F. Nori, Phys. Rev. A
75, 063818 (2007).

\bibitem{sakurai} J. J. Sakurai, \textit{Modern Quantum Mechanics }%
(Addison-Wesley, Reading, Ma, 1994).

\bibitem{landau} L. D. Landau and E. M. Lifshitz, \textit{Quantum Mechanics
(Non-relativistic Theory)} (Butterworth-Heinemann, Oxford, 1977) Third
Edition.

\bibitem{meystre3} D. Meiser and P. Meystre, Phys. Rev. A 74, 065801 (2006)

\bibitem{knight} S. Bose, K. Jacobs, and P. L. Knight, Phys. Rev. A 56, 4175
- 4186 (1997).

\bibitem{quan} H. T. Quan, Z. Song, X. F. Liu, P. Zanardi, and C. P. Sun,
Phys. Rev. Lett. 96, 140604 (2006) .

\bibitem{orszag} M. Orszag, \textit{Quantum Optics: Including Noise
Reduction, TrappedIons, Quantum Trajectories and Decoherence}
(Springer-Verlag, Berlin Heidelberg, 2000).

\bibitem{x x yi} X. X. Yi, H. Y. Sun, and L. C. Wang, arXiv:0807.2703 (2008).
\end{thebibliography}
\end{document}